\documentclass[%
nofootinbib,
 amsmath,amssymb,
 aps,
 prd,
floatfix,notitlepage
]{revtex4-1}
\usepackage[normalem]{ulem}
\usepackage{slashed}
\usepackage{graphicx}
\usepackage{dcolumn}
\usepackage{bm}
\usepackage{hyperref}
\usepackage{orcidlink}
\usepackage{footnote}
\usepackage{hyperref}
\usepackage{xcolor}
\usepackage{subcaption}

\begin{document}

\title{Cooling of quark stars from perturbative QCD}

\author{Úrsula {\sc Fonseca}\orcidlink{??}}
 \affiliation{
 Instituto de F\'\i sica, Universidade Federal do Rio de Janeiro,\\
 CEP 21941-972 Rio de Janeiro, RJ, Brazil 
}

\author{Eduardo S. {\sc Fraga}\orcidlink{0000-0001-5340-156X}}
\affiliation{
 Instituto de F\'\i sica, Universidade Federal do Rio de Janeiro,\\
 CEP 21941-972 Rio de Janeiro, RJ, Brazil 
}

\begin{abstract}
We investigate the thermal evolution of quark stars with and without a hadronic crust using an equation of state derived from perturbative QCD that incorporates the running of the strong coupling and the strange quark mass. Our analysis reveals that bare quark stars cool too rapidly to match the luminosity data, including those of the coldest observed isolated neutron stars, even when the uncertainty from the renormalization scale is taken into account. In contrast, configurations featuring a hadronic crust exhibit slower cooling and improved agreement with observational data. We also observe that the cooling band for bare quark stars narrows significantly after $t \sim 1$ year, whereas the configurations with a crust exhibit a larger uncertainty throughout their time evolution. 
\end{abstract}

\maketitle


\section{Introduction}

Neutron stars are believed to occupy a region of the Quantum Chromodynamics (QCD) phase diagram characterized by low temperatures and high baryon densities. While it may be surprising at first glance to call neutron stars cold, since they are born in supernova explosions with temperatures of $T \sim 10^{11}\ \mathrm{K} \sim 10\ \mathrm{MeV}$ and cool over time to temperatures in the keV range \cite{Schmitt_2010, Burgio:2021vgk}, these values are small compared to the energy scale set by QCD. For example, the temperature of the deconfinement transition at vanishing chemical potential is around $T_c \sim (150-160)\ \mathrm{MeV}$ \cite{Cheng:2009zi, Borsanyi:2010cj, Borsanyi:2013bia, Kanaya:2017cpp}, and the quark chemical potential to be found in the interior of neutron stars is typically within the range of hundreds of MeV \cite{Schaffner-Bielich:2020psc}. Because neutron star temperatures are far below such scales, their matter is usually treated within the framework of cold and dense QCD \cite{Schaffner-Bielich:2020psc,Lovato:2022vgq}.

To describe their structure and dynamical properties, a key ingredient is the equation of state (EoS), which encodes the thermodynamics of strongly interacting matter. In particular, the relevant regime involves baryon number densities around and above nuclear saturation density, $n_0 \approx0.16\ \mathrm{fm}^{-3}$, or $n_0m_u\approx 2.7\times10^{14} \ \mathrm{g}\ \mathrm{cm}^{-3}$. However, the EoS of neutron star matter remains uncertain, since first-principle approaches are currently reliable only in the limits of very low or very high baryon densities.

At low densities -- typically from bellow to slightly above $n_0$ -- quarks are confined within hadrons, and the interactions among them can be described using the first-principle approach of chiral effective field theory ($\chi \mathrm{EFT}$) \cite{Weinberg:1978kz,Epelbaum:2008ga,Piarulli:2019pfq,Drischler_2021}. This density regime corresponds to the outer layers of the star, from the atmosphere down to the inner crust. Current calculations can reach up to baryon densities $\sim 1.1\ n_0$
\cite{Tews:2012fj,Hebeler:2013nza}. 
As one moves deeper into the star, toward its core, baryon density increases significantly, typically reaching values in the range of $2-10 \ n_0$. This intermediate-density regime is the most challenging to describe from first principles. Unfortunately, Lattice QCD simulations cannot be applied in this region due to the Sign Problem \cite{Aarts:2015tyj}. Therefore, one does not have this benchmark to compare to within the entire physical environment of neutron stars.

At much higher densities, around $n\approx 40 n_{0}$, the QCD coupling becomes weak due to asymptotic freedom, making perturbative QCD (pQCD) reliable to compute the equation of state. Although such densities exceed those expected in the core of compact stars, pQCD imposes very stringent constraints to the EoS \cite{Kurkela:2014vha,Fraga:2015xha, Komoltsev:2021jzg, Gorda:2022lsk, Gorda:2023usm}. Besides, it provides a systematic, controlled calculation, grounded in the fundamental theory of strong interactions, with a clear error estimate given by the freedom in the choice of the renormalization scale. 
The equation of state for a system composed of up, down and strange quarks at zero temperature and nonzero baryon chemical potential was originally obtained within perturbative QCD more than four decades ago by Freedman and McLerran \cite{Freedman:1976ub,Freedman:1977gz}, and also by Baluni \cite{Baluni:1977ms} and Toimela \cite{Toimela:1984xy}. Since then, it has been systematically improved \cite{Blaizot:2000fc,Fraga:2001id,Fraga:2004gz,Kurkela:2009gj,Fraga:2013qra,Kurkela:2014vha,Fraga:2015xha,Ghisoiu:2016swa,Annala:2017llu,Gorda:2018gpy,Annala:2019puf,Gorda:2021kme,Gorda:2021znl}.

In this work, we investigate the thermal evolution of quark stars -- hypothetical compact objects composed entirely of deconfined strange quark matter \cite{Bodmer:1971we, Witten:1984rs, Terazawa:1989iw} -- with and without a hadronic crust, using an equation of state derived from perturbative QCD to order $\alpha_s$. We incorporate the renormalization group running of the strong coupling, $\alpha_{s}$, and the strange quark mass, $m_s$. We focus on the thermal emission of these objects. As will be discussed in detail, such stars lose thermal energy via two primary mechanisms: neutrino emission from the bulk and photon emission from the surface \cite{1998PhR...292....1T}. During the early stages of evolution, neutrino emission dominates the cooling and is highly sensitive to the microscopic composition of the star. Among the various neutrino processes, the most efficient is the direct Urca process \cite{Gamow:1940eny,Gamow:1941gis}. In nucleonic matter, this process is only kinematically allowed when the proton fraction exceeds a critical threshold \cite{Lattimer:1991ib, 1999A&A...346..465S, Mendes:2024hbn}. In quark matter, however, the direct Urca process can occur if interactions between quarks are included (i.e., if the strong coupling $\alpha_{s}$ is non-zero) and if electrons are present \cite{Blaschke_2000}.

Observationally, there are at least three young neutron stars whose surface temperatures are too low to be explained by the minimal cooling scenario \cite{Marino:2024gpm}, which excludes fast neutrino processes such as direct Urca. These data provide strong evidence that enhanced neutrino emission mechanisms must be active in at least some compact stars. 

Our analysis reveals that bare quark stars cool too rapidly to match the luminosity data, including those of the coldest observed isolated neutron stars, even when the uncertainty from the renormalization group scale is taken into account. In contrast, configurations featuring a hadronic crust exhibit slower cooling and improved agreement with observational data. We also observe that the cooling band for bare quark stars narrows significantly after $t \sim 1$ year, whereas the configurations with a crust exhibit a larger uncertainty throughout their time evolution. 

The paper is organized as follows. In Section \ref{sec:thermal_ev}, we present and briefly discuss the thermal evolution of compact stars, including the microscopic inputs required to solve the relevant equations. Section \ref{sec:eos} focuses on discussing the equation of state employed in our calculations, including the presence or absence of a hadronic crust. Our main results and their discussion are presented in Section \ref{sec:results}. There we show temperature profiles and the luminosity evolution. The latter is directly compared to current observational data. Finally, Section \ref{sec:summary} summarizes the findings and outlines perspectives for future work.

\section{Thermal evolution}
\label{sec:thermal_ev}

\subsection{Evolution equations}

Approximately one minute after its formation in a supernova explosion, the newly born compact star cools to a temperature bellow $\sim 1\ \mathrm{MeV} \approx 1.16 \times 10^{10} \ \mathrm{K}$ \cite{Schmitt_2010}. At this stage, the star becomes effectively transparent to neutrinos, and we focus our analysis exclusively on this phase. From this point onward, cooling proceeds through two dominant mechanisms: neutrino emission from the entire stellar body and heat conduction from the internal layers, resulting in the thermal emission of photons. These energy loss channels operate simultaneously, with neutrino cooling dominating early evolution and photon emission taking over at later times \cite{Page_2000}. The precise time at which surface photon emission becomes comparable to or dominates over neutrino cooling depends on the properties of the star.
The thermal evolution of a spherically symmetric compact star in general relativity is governed by two coupled differential equations: the energy balance equation and the heat transport equation \cite{1977ApJ...212..825T, Weber:2006ep}.

The energy balance equation gives the time dependence of the temperature due to neutrino losses and outward heat flow via 
\begin{equation}
    \label{eq:thermalBalance}
    \dfrac{\partial T}{\partial t} = - \dfrac{1}{4 \pi r^{2} e^{\Phi}C_{\nu}} \sqrt{1-\dfrac{2Gm}{c^{2}r}} \dfrac{\partial}{\partial r}(e^{2\Phi}L) - \dfrac{\epsilon_{\nu}}{C_{\nu}} e^{\Phi}  \, ,
\end{equation}
where $L(r,t)$ is the luminosity (excluding neutrinos), $\Phi(r)$ is the metric function, $\epsilon_{\nu}(r, T)$ is the neutrino emissivity, $C_{v}(r, T)$ is the specific heat and $T(r, t)$ is the temperature.  The factors $e^{\Phi(r)}$ account for the gravitational redshift. 

The second differential equation links the local luminosity to the temperature gradient and is given by
\begin{equation}
    \label{eq:heatTransp}
    L = - 4\pi \kappa r^{2} \sqrt{1-\dfrac{2Gm}{c^{2}r}} e^{-\Phi} \dfrac{\partial}{\partial r}(T e^{\Phi}) \, ,
\end{equation}
where $\kappa$ is the thermal conductivity. One needs to solve Eqs. (\ref{eq:thermalBalance}) and (\ref{eq:heatTransp}) simultaneously to determine $L(r, t)$ and $T(r,t)$.

In addition to the microscopic quantities mentioned above, solving the cooling equations also requires input from macroscopic quantities such as the radial coordinate $r$, the radial profile of the energy density $\epsilon(r)$ and enclosed mass $m(r)$. These are obtained by solving the Tolman-Oppenheimer-Volkoff (TOV) equations for a chosen central density, which requires specifying an equation of state. Since the internal structure of a compact star can be regarded as temperature independent, the TOV equations need to be solved only once.

To solve the thermal evolution equations, appropriate boundary conditions are required. At the center of the star we fix $L(r=0, t)\equiv 0$, which reflects the fact that there is no net energy generation or loss at this point. The outer boundary condition is given by the photon luminosity, which is expressed as
\begin{equation}
    \label{eq:photonLum}
    L_{\gamma}\equiv4\pi \sigma_{SB} T_{e}^{4} \, ,
\end{equation}
where $\sigma_{SB}$ is the Stefan-Boltzmann constant. The quantity $T_e$ denotes the effective temperature at the stellar surface. Its determination, based on the $T_{b}-T_{e}$ relation, will be described in the following paragraph. 

To facilitate numerical computations of the time evolution of the internal temperature and luminosity profiles, it is common to artificially subdivide the problem. This is done by separating the star in two regions: the interior ($r<R_{b}$) and the outer heat-blanketing envelope ($R_{b} \leq r \leq R_{\rm star}$). The standard approach is to define the envelope as the region that extends from the surface of the star to a density of $\rho_{b}=10^{10} \ \mathrm{g}\ \mathrm{cm}^{-3}$, which corresponds to a radial depth of approximately $100\ \mathrm{meters}$ below the surface. This envelope is thin compared to the total radius of the star, and is therefore analyzed separately by solving Eqs. (\ref{eq:thermalBalance}) and (\ref{eq:heatTransp}) under the stationary plane-parallel approximation \cite{Page:2004wb, Page:2005fq}. The solution of these equations provides the relation between the surface temperature, $T_{s} \equiv T_{e}$, and the boundary temperature $T_{b}$. Thus, for a given $T_{b} - T_{e}$ relation, we can express Eq. (\ref{eq:photonLum}) in terms of the boundary temperature. This yields the outer boundary condition, $L(R_{b}) = L_{\gamma}$, used in the cooling calculations. For bare quark stars, we take $T_{e}=T_{b}$; for stars with a crust, following \cite{Horvath:1991ms}, we set $T_{e}=5\times 10^{-2}\ T_{b}$.

The local quantities $L$, $R$, and $T_{e}$ are measured at the stellar surface, whereas a distant observer detects their red-shifted counterparts: $L^{\infty} = e^{2\Phi}L$, $T^{\infty}_{e}=e^{\Phi}T_{e}$, and $R^{\infty}  = e^{-\Phi}R$. The main goal is to determine how the apparent temperature $T^{\infty}$ (or, equivalently, of $L^{\infty}$) evolves with stellar age $t$.

\subsection{Microscopic inputs}
\label{subsec:micInputs}

The thermal evolution depends sensitively on microphysical quantities that control energy loss, heat storage, and thermal transport in the stellar interior \cite{Potekhin:2015qsa}. These, in turn, are functions of the local composition, density, and temperature, and are also influenced by the presence of exotic phases or superfluidity. For this reason, we need to separate the inputs for the core and the crust. In the bare scenario, however, we apply the quark core inputs throughout the entire stellar profile. 

For the neutrino emissivity $\epsilon_{\nu}$ -- which represents the energy radiated per unit volume per unit time via neutrinos -- we include, 
in the quark core, contributions from direct Urca, modified Urca, and bremsstrahlung processes, using the values computed in Ref. \cite{Iwamoto:1982zz}. For the hadronic crust, we consider contributions from plasmon decay and neutrino bremsstrahlung resulting from electron-nucleus collisions. For plasmon decay we implement the interpolation formula from Ref. \cite{Yakovlev:2000jp} and for bremsstrahlung we use the fit provided by Ref. \cite{kaminker1998neutrinopairbremsstrahlungelectronsneutron}. 

The specific heat of strange quark matter is given by the sum of contributions from electrons and quarks, for which we use the expression given in Ref. \cite{Iwamoto:1982zz}. In the crust we must account for the specific heat of electrons, using the same expression as in the core, as well as that of the ions. The ion contribution depends on the properties of the ions present in the crust, such as their Fermi momentum, mass number $A$, and atomic number $Z$. It is also necessary to account for the melting temperature, $T_m$, dependence on these quantities, since different compositions can lead to different melting temperatures. One can have, e.g., a Coulomb crystal at the bottom of the outer crust and a Coulomb liquid at the top of the inner crust \cite{Haensel:2007yy}. This, in turn, results in changes in its thermal properties. 

We adopt the one-component plasma (OCP) approximation. For temperatures below the melting temperature, the specific heat is taken as the sum of the Coulomb crystal contribution in the harmonic approximation, obtained from Ref. \cite{Baiko_2001}, and the anharmonic correction from Ref. \cite{Farouki1993}, both calculated using parameters appropriate for a BCC lattice. For temperatures above the melting point, we compute the specific heat for an ideal ion gas, including the excess terms from the internal energy as given in Refs. \cite{DeWitt1996, Haensel:2007yy}.

Finally, we use the expression given in Ref. \cite{Haensel:1991pi} for the thermal conductivity of strange quark matter. In the crust, the thermal conductivity is conveniently expressed in terms of the effective electron collision frequency, $\nu_{\kappa}$, as \cite{urpin1980thermogalvanomagnetic, ziman2001electrons}
\begin{equation}
\label{eq:thermalConducCrust}
    \kappa = \dfrac{\pi^{2}k_{B}Tn_{e}}{3m_{e}^{*} \nu_{\kappa}} ,
\end{equation}
where $m_{e}^{*} \equiv \epsilon_{F}/c^{2}$ is the effective dynamical mass of an electron at the Fermi surface, $n_{e}$ is the electron number density, $k_{B}$ is the Boltzmann constant, and $\nu_\kappa$ is the effective electron collision frequency, obtained as the sum of the partial collision frequencies corresponding to the relevant electron scattering mechanisms, which can be treated separately \cite{Potekhin:1999yv}.

In our computation, we use the electron transport coefficients provided in Ref. \cite{Potekhin:1999yv}, whose corresponding Fortran code is available online (\url{https://www.ioffe.ru/astro/conduct/index.html}). Their results were computed for the outer envelope of neutron stars. For our calculations, we extrapolate them to the densities found at the bottom of the thin hadronic crust. As shown in Ref. \cite{Gnedin:2000me}, this extrapolation is reasonable within the density regime we consider. However, it no longer holds in the inner crust of neutron stars, where densities exceed the neutron drip threshold, and the presence of free neutrons alters the functional behavior.

\section{Equation of state}
\label{sec:eos}

\subsection{Quark matter}
\label{subsec:QM}

To solve the cooling equations, we need the inputs discussed in \ref{subsec:micInputs} as well as a specification of the equation of state. A commonly adopted equation of state for strange quark matter in the compact star community is the MIT bag model \cite{PhysRevD.9.3471, PhysRevD.10.2599, Johnson:1975zp}, which treats quarks as a degenerate Fermi gas confined to a finite region of space. A phenomenological treatment of confinement is implemented via an additive constant $B$ to the energy of the system, reflecting the difference between the perturbative vacuum and the physical, chiral-symmetry-breaking ground state of the theory. 
Within the framework of the bag model, a standard approach is to use the thermodynamic potential computed in \cite{Farhi:1984qu}, which includes first-order corrections in the strong coupling constant $\alpha_{s}$ from perturbative QCD and accounts for a finite strange quark mass $m_{s}$. In the massless case, these first-order corrections cancel out in the equation of state. 

The resulting equation of state depends on some free parameters that must be specified. For instance, the renormalization scale needs to be chosen to match the typical chemical potentials relevant to the system \cite{glendenning1997compact}. In addition, the values of $\alpha_s$, $m_s$, and $B$ are fixed and must all be specified. A common way to constrain this parameter space is by imposing strange matter stability criteria: the energy per baryon of three-flavor quark matter should be lower than that of iron nuclei, while the corresponding value for two-flavor quark matter must remain higher. 

However, Lattice simulations of hot QCD thermodynamics clearly demonstrate the importance of accounting for the interactions of quarks and gluons to address the phase structure of strongly interacting matter. As discussed in Ref. \cite{Fraga:2013qra}, the bag model description exhibits a very poor behavior with this respect by construction. There is no physical reason to expect a different outcome for cold and dense QCD. Therefore, here we adopt the simplest version of a pQCD equation of state for cold quark matter \cite{Fraga:2004gz}: two massless (up and down) and one massive (strange) quarks, up to $O(\alpha_s)$, using the $\overline{\mathrm{MS}}$ scheme. Renormalization group running effects are included for both the coupling and the strange quark mass. 
The resulting thermodynamic potential depends explicitly on the quark chemical potentials $\mu_i$ ($i=u, d, s$) and the renormalization subtraction point $\bar{\Lambda}$, and implicitly through the scale dependence of the strong coupling $\alpha_s(\bar{\Lambda})$ and the running mass $m_s(\bar{\Lambda}$). 

The only free parameter in the EoS is $\bar{\Lambda}$. Following the discussion of Refs. \cite{Fraga:2001id,Kurkela:2009gj,Fraga:2004gz}, we take $\bar{\Lambda} \propto  \sum{\mu_{i}}/N_{f}$, where $N_f$ is the number of flavors. The canonical value for the proportionality constant is $2$, but it is usually varied between $1$ and $4$ to assess the sensitivity of the results to this scale choice, which provides a band of theoretical uncertainty associated with the equation of state. 
To this order in perturbation theory, however, we find that the equation of state becomes too soft at $\bar{\Lambda} = \mu$ to support stable stellar configurations with masses of $1.4\ \mathrm{M}_\odot$ or greater. Since we are primarily interested in configurations at or above this mass threshold we choose, phenomenologically, to set the lower limit of $\bar{\Lambda}$ to $2\mu$.

In order to obtain electrically neutral quark matter, we include electrons with chemical potential $\mu_e$. This imposes the following condition on the number density $n_i$ of the particles:
\begin{equation}
\label{eq:chargeNeu}
\sum_{i=u, d, s, e} q_{i} n_{i} = \dfrac{2}{3}n_u - \dfrac{1}{3}n_d - \dfrac{1}{3}n_s - n_e = 0 \,.
\end{equation}

\begin{figure}[htbp]
    \centering
    \begin{minipage}[b]{0.45\textwidth}
        \centering
        \includegraphics[width=\textwidth]{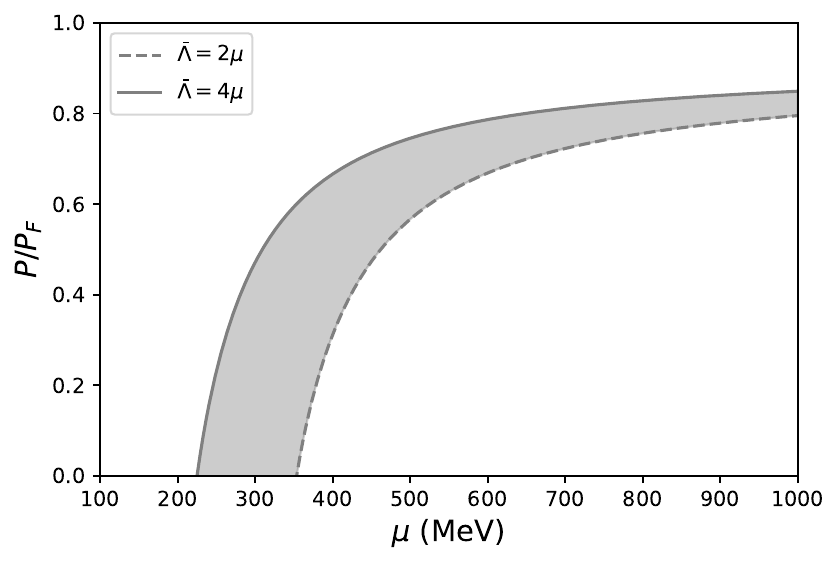}
        \label{fig:ppf}
    \end{minipage}
    \hfill
    \begin{minipage}[b]{0.45\textwidth}
        \centering
        \includegraphics[width=\textwidth]{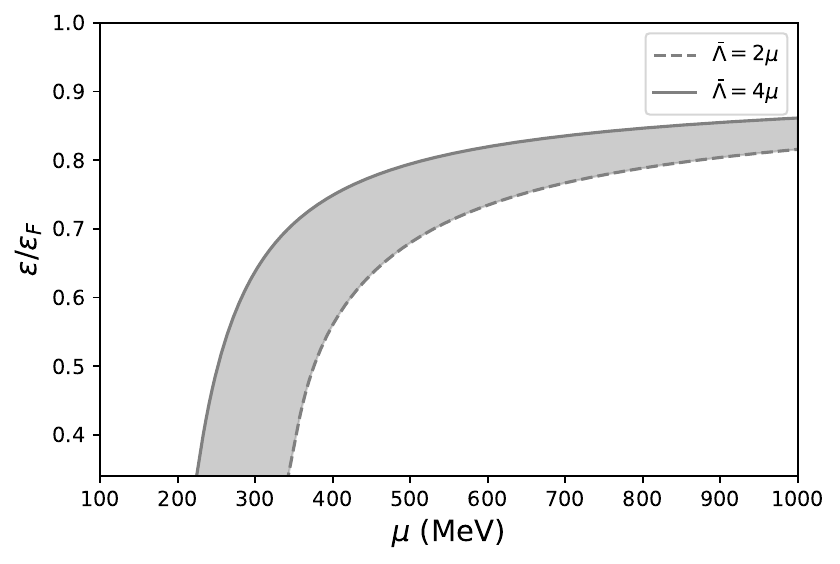}
        \label{fig:eef}
    \end{minipage}
    \caption{\raggedright Pressure (left panel) and energy density (right panel), each scaled by their respective Fermi values, shown as functions of the strange quark chemical potential $\mu$. Bands represent the renormalization-scale
dependence in the range between $\bar{\Lambda} = 2\mu$ and $\bar{\Lambda} = 4\mu$. }
    \label{fig:ppfeef}
\end{figure}

We also impose beta equilibrium conditions, the relevant processes being
\begin{equation}
\begin{aligned}
&d \rightarrow u + e + \bar{\nu}_e \, , &&u + e \rightarrow d + \nu_e \, , \\
&s \rightarrow u + e + \bar{\nu}_e \, , &&u + e \rightarrow s + \nu_e \, , \\
&&s + u \leftrightarrow d + u \, ,
\end{aligned}
\end{equation}
which yield the following conditions for the quark and electron chemical potentials:
\begin{equation}
\begin{aligned}
&\mu_d = \mu_s = \mu \, , &&\mu_u + \mu_e = \mu \, .
\end{aligned}
\end{equation}
Here, we have set the neutrino chemical potential to zero, since neutrinos escape from the star rapidly.
With these two conditions, we ensure that only one independent chemical potential remains, which we take to be $\mu_s = \mu$. This allows us to express the thermodynamic potential solely as a function of $\mu$. Consequently, the equation of state can be obtained from the parametrized expressions for pressure and energy density as functions of $\mu$. In Figure \ref{fig:ppfeef}, we show the pressure and energy density as functions of $\mu$, normalized by the free (massless) Fermi gas result, computed for $\bar{\Lambda} = 2\mu$ and $\bar{\Lambda} = 4\mu$. Both panels exhibit a notable increase in the band width at lower quark chemical potentials, as expected.

\subsection{Crust}
\label{subsec:crust}

Strange quark stars may be surrounded by a thin crust of hadronic matter, suspended above the core by electrostatic forces. It was demonstrated that the electrons that neutralize the positively charged strange matter are bound primarily by the Coulomb interaction and extend several hundred fermis beyond the quark matter surface \cite{Alcock1986, AlcockOlinto1988, Kettner1995}. This spatial separation results in the formation of a surface electric dipole layer composed of a positively charged layer just beneath the surface and the extended electron cloud above it. The strong electric field generated in this region is capable of supporting a crust of nuclear material physically detached from the core.

The resulting vacuum gap between the quark core and the crust is estimated to span several hundred fermis \cite{1992ApJ...400..647G, Stejner_2005}. The maximum density at the base of the crust is generally constrained by the neutron drip density, $\rho_{\mathrm{drip}} \sim4\times10^{10}\ \mathrm{g}\ \mathrm{cm}^{-3}$. Beyond this threshold, free neutrons would no longer be bound in nuclei, allowing them to gravitationally sink into the core. Upon contact with the strange matter, these neutrons would be rapidly converted into quark matter, thereby destabilizing the crust \cite{1995ApJ...450..253G}. 

To include this crust in our calculations, we adopt the Baym–Pethick–Sutherland (BPS) EoS \cite{Baym:1971pw}, which describes matter in its cold, catalyzed state, and match it to the quark matter EoS. This results in an overall EoS composed of two segments, one for the nuclear crust and another for the quark matter core. Since the only constraint on the crust is its maximum density, quark stars could in principle possess crusts of varying thicknesses, depending on their formation history and age, with the only requirement being that the transition occurs at or below $\rho_{\mathrm{drip}}$. The resulting equation of state is represented by

\begin{equation}
    P(\epsilon) = 
    \begin{cases}
    P_{\mathrm{BPS}}(\epsilon), & \text{if } P < P_{\mathrm{crust}} \\
    P_{\mathrm{pQCD}}(\epsilon), & \text{if } P \geq P_{\mathrm{crust}}.
    \end{cases}
\end{equation}

\begin{figure}[h]
    \centering
    \includegraphics[height=7cm]{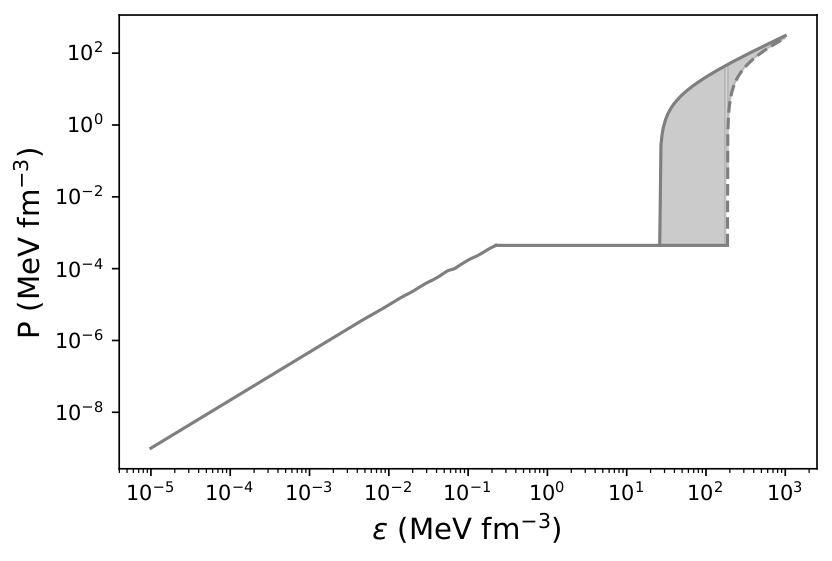}
    \caption{\raggedright Equation of state $P(\epsilon)$ for a quark star with a nuclear crust. The low-density sector is described by a BPS equation of state, whereas the high-density sector is built from cold and dense pQCD.}
    \label{fig:eos_completa}
\end{figure}

In all configurations, the surface of the star is defined by the condition $P=0$. However, when a crust is present, the edge of the quark core is located at a nonzero pressure, equal to the pressure of the crustal material at the core–crust transition density, i.e., $P=P_{\mathrm{crust}}$. Therefore, the presence of a crust slightly compresses the quark core compared to the bare case.

In this work, we adopt a maximally thick crust, with the crust–core transition taking place at $\rho_{\mathrm{drip}} \approx 4 \times 10^{11}\ \mathrm{g}\ \mathrm{cm^{-3}}$. Although this maximum density estimate was later revised \cite{1997A&A...325..189H}, indicating a lower crustal density, we proceed with the conventional maximally thick-crust assumption to provide an upper-limit scenario for the crustal contribution. The resulting equation of state is shown in Figure \ref{fig:eos_completa}.
The low-density regime is described by the BPS EoS. This ceases to be valid at the neutron drip point with $\epsilon \approx 0.22\ \mathrm{MeV}\ \mathrm{fm}^{-3}$ and $P_{\rm drip}\approx 4.4\times 10^{-4}\ \mathrm{MeV}\ \mathrm{fm}^{-3}$. For the core, we employ the pQCD EoS, starting at $P_{\rm drip}$, evaluated for the two limiting renormalization scales $\bar{\Lambda} = 2\mu$ and $\bar{\Lambda} = 4\mu$, using the same line-style convention as in Figure \ref{fig:ppfeef}. The two EoS are matched at $P_{\rm drip}$, corresponding to $\epsilon \approx 26\ \mathrm{MeV}\ \mathrm{fm}^{-3}$ for $\bar{\Lambda}=2\mu$ and $\epsilon \approx 186\ \mathrm{MeV}\ \mathrm{fm}^{-3}$ for $\bar{\Lambda}=4\mu$. The discontinuity at the junction reflects the absence of a mixed phase in the intermediate-density region.

\section{Results}
\label{sec:results}
\begin{figure}[h]
    \centering
    \includegraphics[height=7cm]{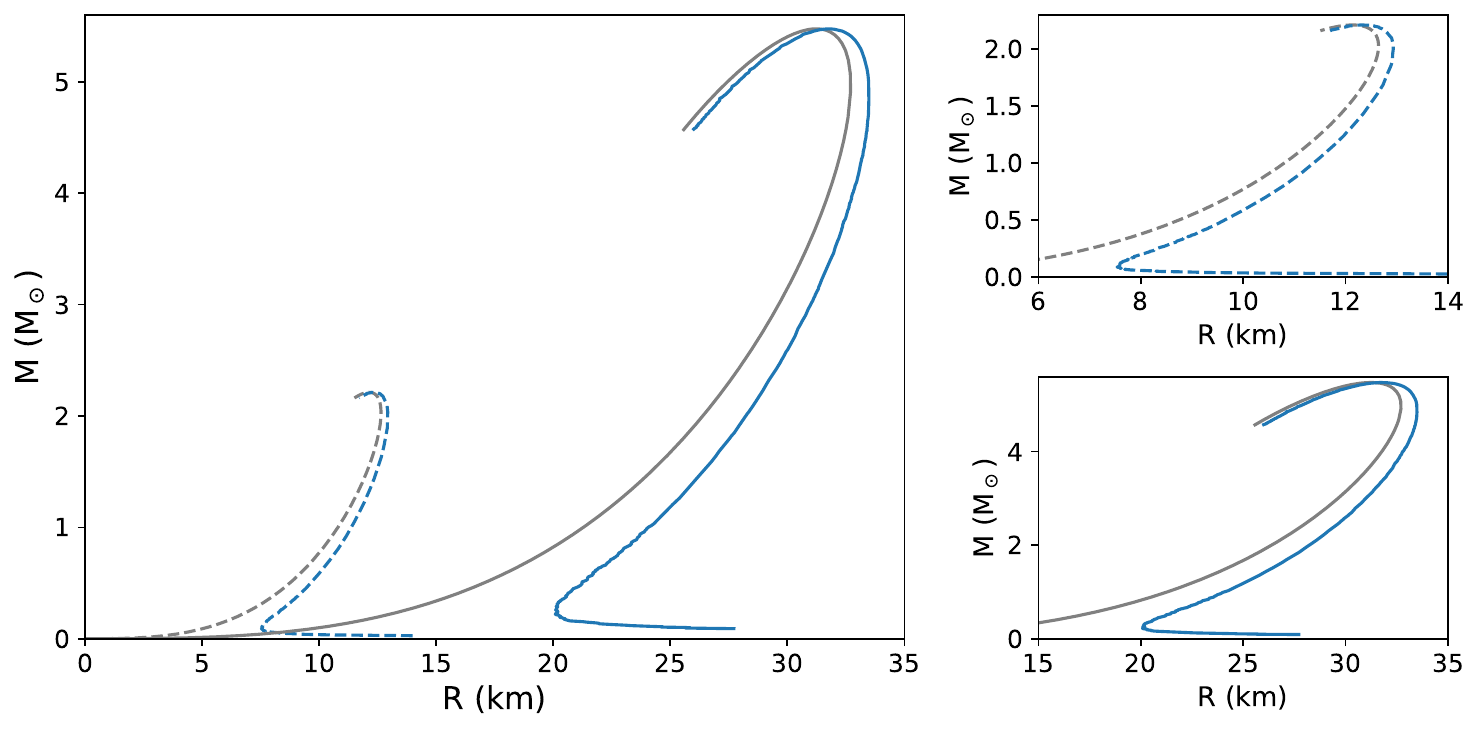}
    \caption{\raggedright The main (left) panel shows the full mass-radius diagram computed from the pQCD EoS. Gray curves correspond to configurations without a crust, whereas blue curves include a nuclear crust. The right-hand panels provide zoom views isolating the impact of the crust for the cases $\bar{\Lambda}=2\mu$ (top) and $\bar{\Lambda}=4\mu$ (bottom). }
    \label{fig:mr}
\end{figure}

We show in Figure \ref{fig:mr} the mass-radius relations resulting from the numerical solutions of the Tolman-Oppenheimer-Volkoff (TOV) equations \cite{Tolman:1939jz, Oppenheimer:1939ne}, which describe the hydrostatic equilibrium of a spherically symmetric object within the framework of General Relativity. Unlike the thermal evolution equations, the only microscopic input required to solve the structure equations is the EoS. The resulting curves depict the stellar mass and radius obtained for different values of central pressure $p(r=0)$ using the pQCD EoS described in Section \ref{sec:eos} with running strong coupling $\alpha_s$ and running strange quark mass $m_s$. Curves corresponding to quark stars without a crust are shown in gray, whereas those including a hadronic crust are shown in blue. 

As expected, including a crust removes the self-bound nature of quark stars and modifies the stellar radius as a function of mass. For very low-mass strange quark stars, the crust can extend several kilometers beyond the radius of the bare configuration, whereas for more massive stars it becomes much thinner. Although the crust noticeably alters the stellar radius, the total mass remains essentially unchanged, with differences on the order of $10^{-5} M_\odot$, which are negligible at the scale of the plotted results. 
In general, configurations (with and without a crust) corresponding to $\bar{\Lambda} = 4\mu$ are less compact than those with $\bar{\Lambda} = 2\mu$. Moreover, the impact of the crust on the radius is more pronounced in the $\bar{\Lambda} = 4\mu$ case, indicating a comparatively thicker crust for these configurations.

In what follows, we compute cooling curves, i.e., the evolution of $L_\infty$ with stellar age, for quark stars with and without a crust, considering configurations with mass $M = 1.4 M_\odot$. The results obtained with the pQCD equation of state are shown in gray, where the cases with $\bar{\Lambda} = 2\mu$ (dashed line) and $\bar{\Lambda} = 4\mu$ (solid line) define the lower and upper limits of the gray band. For comparison, we also compute the cooling curve using the MIT bag model with $\mathcal{O}(\alpha_s)$ corrections \cite{Farhi:1984qu} in which the parameters are set to $\alpha_s=0.4$, $m_s=100\ \mathrm{MeV}$ and $B^{1/4}=128.9\ \mathrm{MeV}$. The parameters were chosen such that the central density of the resulting $1.4\ \mathrm{M}_{\odot}$ stellar configuration lies within the range spanned by the central densities of the corresponding $1.4\ \mathrm{M}_{\odot}$ stars obtained from the upper and lower pQCD band limits. This result is shown in black in the cooling curves.

For all results presented here, we assume that the initial configuration of the star is a constant redshifted temperature profile, $\tilde{T}(r) = Te^{\Phi(r)}$, and we fix the initial temperature $T = 10^{9}\ \mathrm{K}$ for both bounds of the band across all configurations. Previous analyses \cite{1991ApJS...75..449V, Schaab:1996jm} have shown that the choice of the initial temperature profile does not significantly affect the resulting cooling curves. Therefore, we adopt this simple initial condition without the need to estimate a more detailed temperature distribution. 

We compare our theoretical results with observational data from the catalogue of thermally emitting, middle‑aged isolated neutron stars (INSs), first presented in \cite{Potekhin_2020} and maintained as a continuously updated online resource (\url{https://www.ioffe.ru/astro/NSG/thermal/index.html}), which provides their ages and thermal luminosities. The data is divided into four classes:

\begin{itemize}
    \item \textit{Weakly magnetized thermal emitters (CCO-like)}: with inferred or upper-limit dipolar fields $B_\mathrm{dip} \lesssim 5\times10^{11}\,\mathrm{G}$, or undetermined field strengths (depicted in red).
    
    \item \textit{Ordinary rotation-powered pulsars}: with typical dipolar fields $B \sim 10^{12}\text{--}10^{13}\,\mathrm{G}$ (depicted in blue).
    
    \item \textit{High-$B$ pulsars}: with inferred $B \sim 10^{13}\text{--}10^{14}\,\mathrm{G}$ (depicted in green).
    
    \item \textit{Small-area thermal emitters}: group of pulsars with very small effective thermally emitting areas ($R_\mathrm{eff}^\infty \lesssim 0.5\,\mathrm{km}$), whose thermal luminosities can be regarded as upper limits to the total surface emission (depicted in purple).
\end{itemize}

We restrict our sample to neutron stars whose ages have been determined independently of timing measurements. These ages are estimated from the proper motion of the star, physical properties of the associated supernova remnant or surrounding nebula or, in a few cases, on historical supernova dates. 

\begin{figure}[h]
    \centering
    \includegraphics[height=7cm]{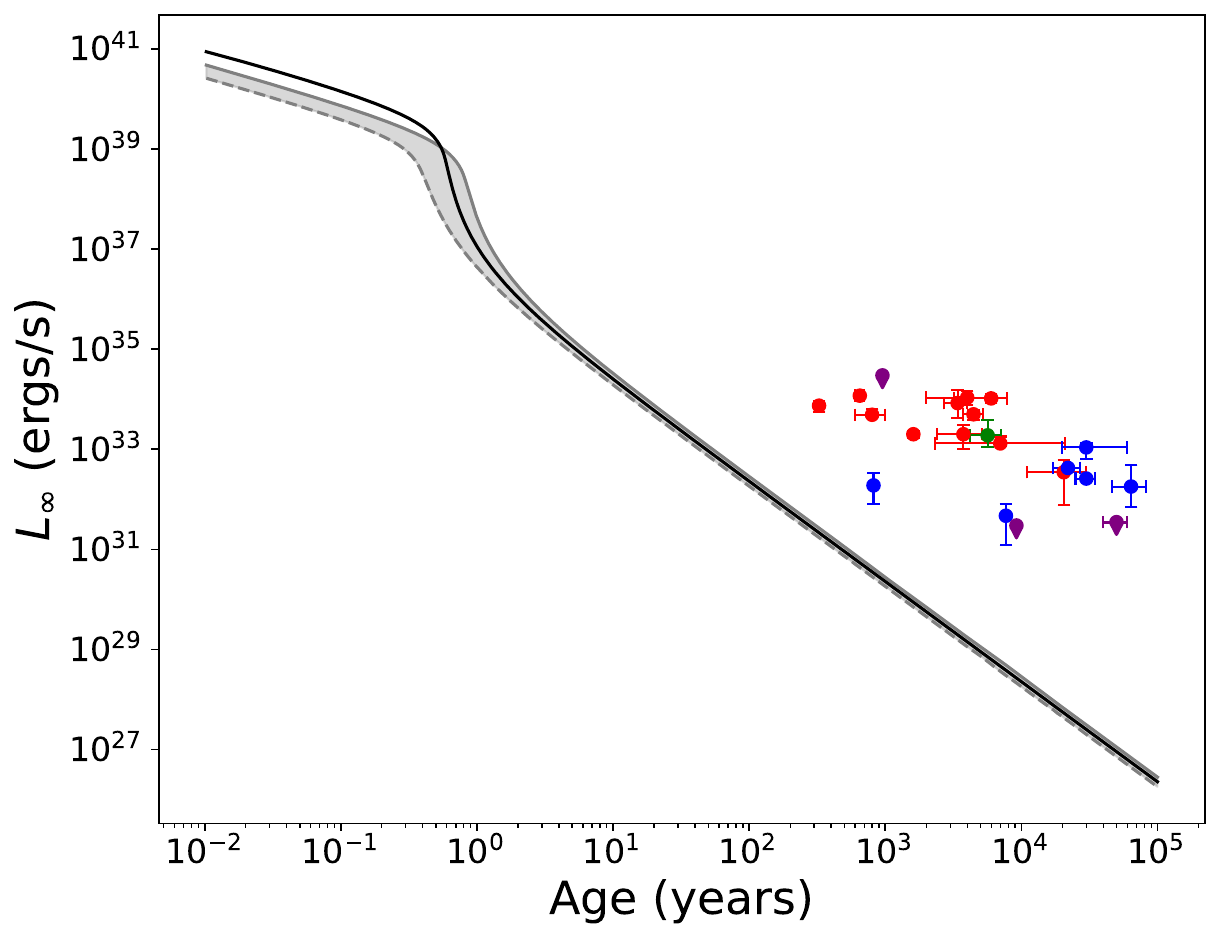} 
    \caption{\raggedright Redshifted photon luminosity $L_{\infty}$ as a function of stellar age for a quark star with $M=1.4\ M_{\odot}$. Observed data points correspond to measured thermal emission from isolated neutron stars (see main text for details). }
    \label{fig:bare_running}
\end{figure}

In Figure \ref{fig:bare_running} we show the resulting cooling curve for a quark star without a crust and compare it to the available observational data. This result is consistent with previous work \cite{Page_2000}, which shows that in bare quark stars the direct Urca process, operating throughout the entire star, leads to extremely efficient cooling, provided that no suppression mechanism -- such as the pairing gaps induced by color superconductivity -- is present. However, several studies on the thermal evolution of quark stars and hybrid stars with quark cores have shown that color-superconducting phases can strongly suppress the direct Urca process \cite{2001ARNPS..51..131A, 2008RvMP...80.1455A}, leading to significant changes in the cooling behavior \cite{Hess:2011qw, Negreiros_2012, Zapata_2022}

As a consequence of the absence of suppression mechanisms, the star reaches very low luminosities at early ages, becoming too cold to account for the observational data, even for the coldest observed isolated neutron stars. We also observe in the figure that the cooling band that measures the uncertainty from perturbative QCD begins to narrow significantly within the first year of evolution. When the star reaches two years of evolution, the upper and lower limits begin to exhibit similar behavior, and the band width remains constant till the end of our analysis. 

\begin{figure}[h]
    \centering
    \includegraphics[height=7cm]{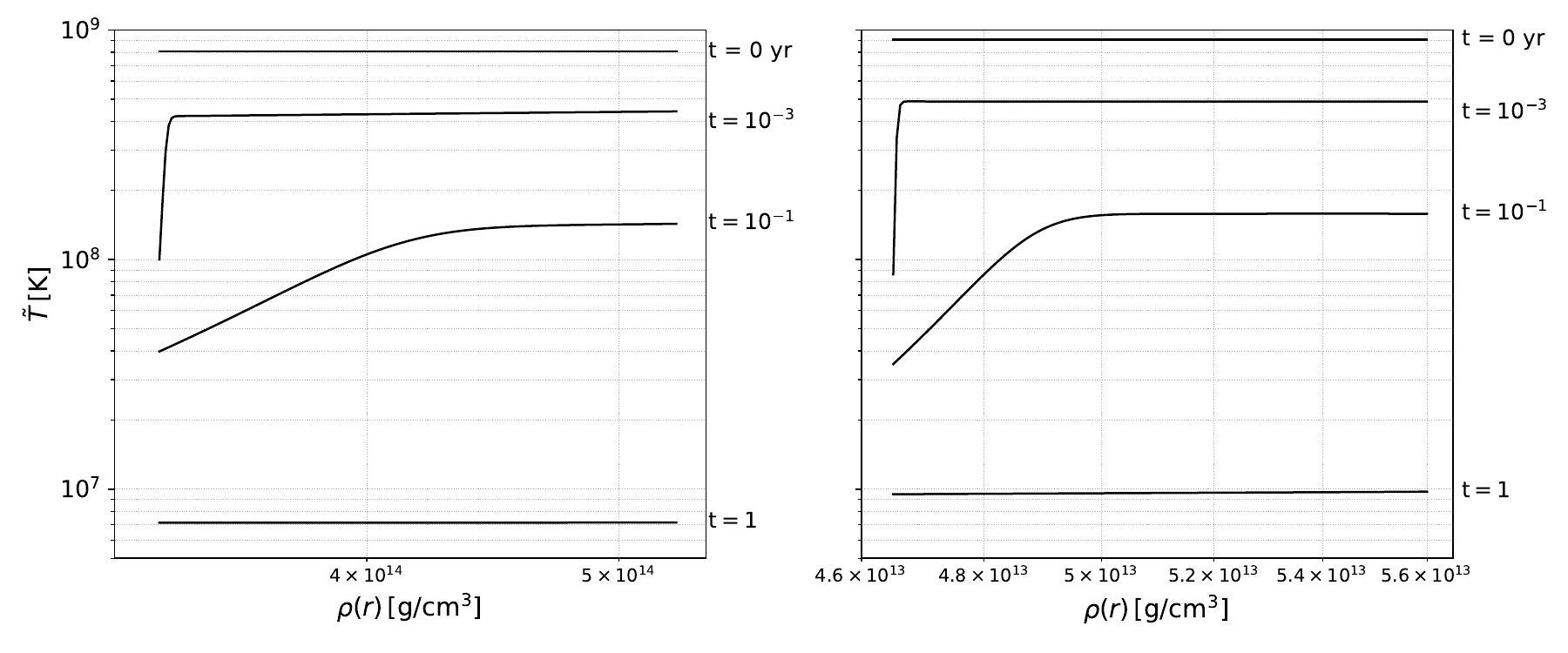}
    \caption{\raggedright Radial temperature profiles of a $1.4\ M_{\odot}$ bare quark star during the early thermal evolution, computed for different stellar ages $t$, using pQCD EoS with $\bar{\Lambda}=2\mu$ (left) and $\bar{\Lambda}=4\mu$ (right). Each curve is labeled by its corresponding age in years, from $t=0$ (initial condition) to $t=1 $ yr.}
    \label{fig:bare_Tprofile}
\end{figure}

To better understand the behavior observed in the cooling band, we analyzed the evolution of the internal temperature at the band boundaries during the early stages of evolution, specifically up to the first year, where the uncertainty is larger in comparison to the rest of the evolution. The results are presented in Figure \ref{fig:bare_Tprofile}. The left panel shows the temperature evolution at the lower bound $\bar{\Lambda}=2\mu$, while the right panel corresponds to the upper bound $\bar{\Lambda}=4\mu$.

The subtle difference in the initial conditions between the two limits arises from the redshift factor $e^{\Phi(r)}$, which differs between the configurations due to their different metric functions. In both cases, the initial temperature drop occurs in the outer layers, where thermal photon emission -- used as a boundary condition in the cooling equations -- takes place, while the core remains relatively hotter.  By $t=10^{-3}$ yr, the temperature profiles are quite similar across both configurations, with some differences: the upper limit exhibits a slightly higher temperature throughout most of the stellar profile compared to the lower limit, on the order of $10^{7} \ \mathrm{K}$. Moreover, in contrast to the interior, the surface temperature is slightly lower by a comparable amount. 

We observe a similar behavior at $t=10^{-1}$ yr, with a hotter interior and a steeper temperature gradient in the outer layers. However, at this point we begin to access the differences in the thermalization process between the limits considered. The configuration with $\bar{\Lambda}=4\mu$ appears to thermalize more rapidly than that with $\bar{\Lambda}=2\mu$, since it exhibits a more extended region within the star that is isothermal with the center. 
By the time both configurations reach $\mathrm{t}=1$ yr, the star becomes nearly isothermal, with a temperature difference between the center of the star ($r=0$) and the surface ($r=R_{\rm star}$) of the order of $10^{4}\ \mathrm{K}$ and $10^{5}\ \mathrm{K}$ for the lower and upper bounds, respectively. This difference decreases further to the order of $10^{0}$ K at $t=14$ yr for $\bar{\Lambda} = 2\mu$ and $t=17$ yr for $\bar{\Lambda} = 4\mu$.

\begin{figure}[h]
    \centering
    \includegraphics[height=7cm]{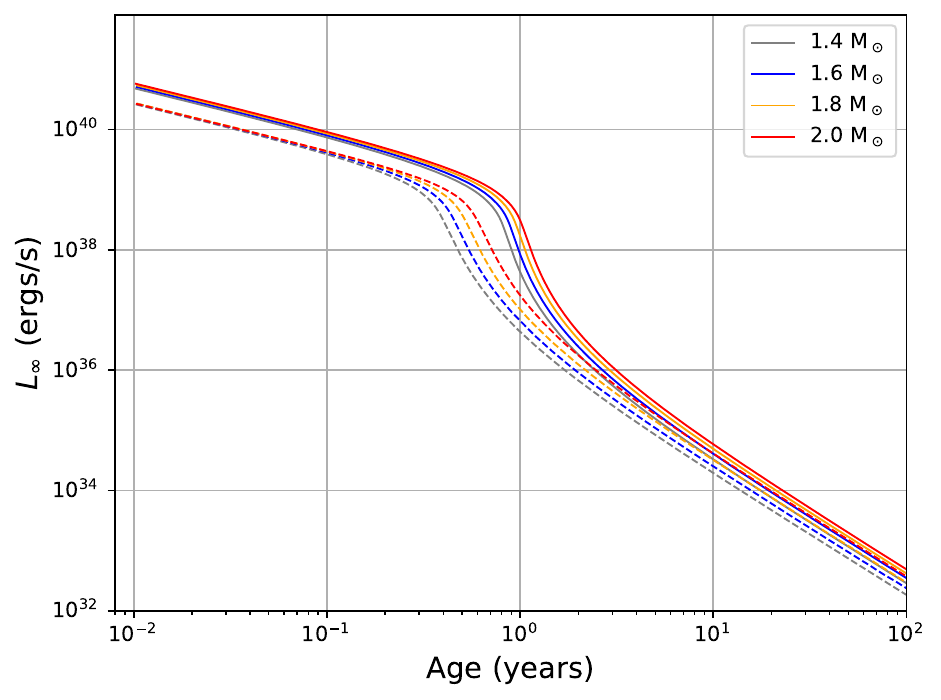}
    \caption{\raggedright Redshifted photon luminosity $L_{\infty}$ as a function of stellar age for quark stars with different masses. }
    \label{fig:bare_masses}
\end{figure}

We also analyze the thermal evolution of quark stars without a crust across a range of stellar masses, as shown in Figure \ref{fig:bare_masses}. The results indicate that increasing the stellar mass systematically shifts the cooling bands toward later times, corresponding to higher surface luminosities at a given age. Moreover, the temporal evolution of the width of the cooling band also exhibits a mass dependence. For more massive configurations, the band is slightly broader during the early stages of evolution and becomes narrower -- relative to the $1.4\ \mathrm{M}_{\odot}$ case -- after the point of the steepest luminosity decline. This narrowing is primarily due to the lower boundary of the band shifting more significantly with increasing mass than the upper boundary, leading to a reduced uncertainty in this evolutionary phase for higher-mass stars.

For quark stars featuring a hadronic crust, we compute the thermal evolution using two distinct numerical approaches, which we refer to as Grid Model A and Grid Model B. Grid Model A explicitly accounts for the presence of the gap between the quark core and the crust. In this case, when constructing the radial grid for the calculations, we place a grid point at the location of the gap by employing a sufficiently small radial step size ($dr$). Grid Model B, in contrast, neglects the gap entirely, initiating the crustal grid at the neutron drip density and continuing outward, down to a density of $\rho = 10^{10}~\mathrm{g \, cm^{-3}}$. 

While both approaches yield consistent results in the region of the cooling curve where observational data exist, they differ at early times: the explicit treatment of the gap affects the initial luminosity evolution and the early internal temperature profiles. Since this young epoch is not constrained by observations, and because resolving the narrow gap--of order several hundred femtometres--significantly increases computational cost, we adopt the second method when comparing our results with observational cooling data (including the comparison with the MIT bag model). Nevertheless, for completeness, we will explicitly show the differences between the two approaches by presenting the pQCD luminosity evolution and the corresponding internal temperature profiles obtained from both methods.

\begin{figure}[h]
    \centering
    \includegraphics[height=7cm]{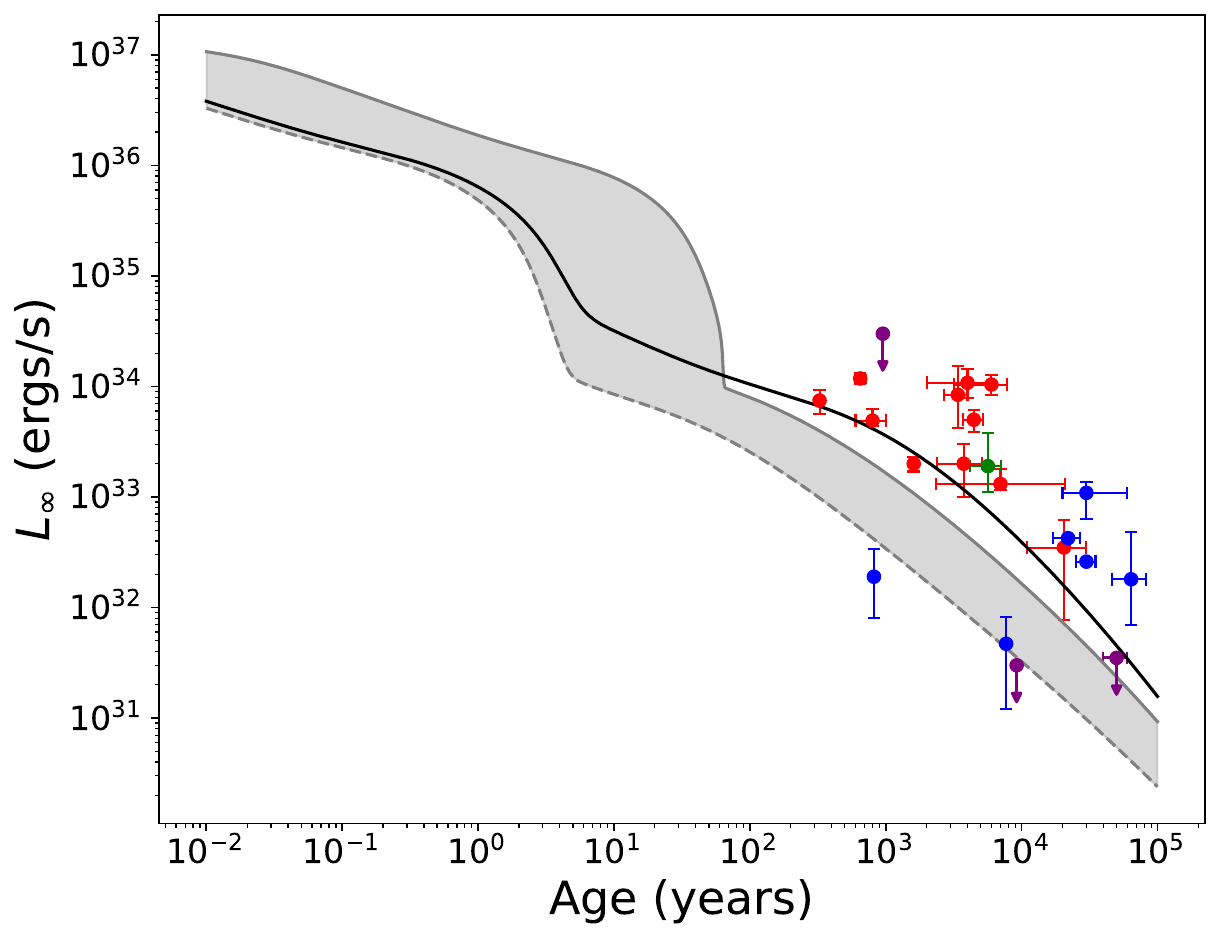}
    \caption{\raggedright Redshifted photon luminosity $L_{\infty}$ as a function of stellar age for a $M=1.4\ M_{\odot}$ quark star with a hadronic crust. }
    \label{cooling_crust}
\end{figure}

In Figure \ref{cooling_crust} we show the resulting cooling band for the $M=1.4\ \mathrm{M}_{\odot}$ quark star with a hadronic crust. 
At early times ($t=10^{-2}$ yr), the luminosity band begins at lower values compared to the case of a bare quark star. Given that the initial temperature is chosen to be $T=10^{9}\ \mathrm{K}$ for all configurations analyzed in this work, this reduction in surface photon luminosity is attributed to the $T_{b}-T_{e}$ relation, which modifies the outer boundary condition for calculations in the presence of a crust. Although the crusted model exhibits lower initial luminosity, the following evolution does not undergo the same sharp decline observed in the bare star scenario. As a result, the crusted configurations remain comparatively warmer at later times for the same stellar age, producing cooling curves that move closer to the observational data.

Furthermore, the luminosity band of the crusted quark star configurations remains systematically broader than that of the bare case at all ages. As in the bare quark star scenario, the widening of the band is associated with the stellar age at which the luminosity curve exhibits its steepest decline. Since the upper and lower boundaries reach this point at distinct stellar ages, the luminosity spread is consequently stretched during this phase. 

This behavior signals that the sensitivity to the choice of the renormalization scale becomes more pronounced in the cooling of quark stars with a hadronic crust. 

\begin{figure}[htbp]
    \centering
    \begin{subfigure}[b]{\linewidth}
        \centering
        \includegraphics[width=0.8\textwidth]{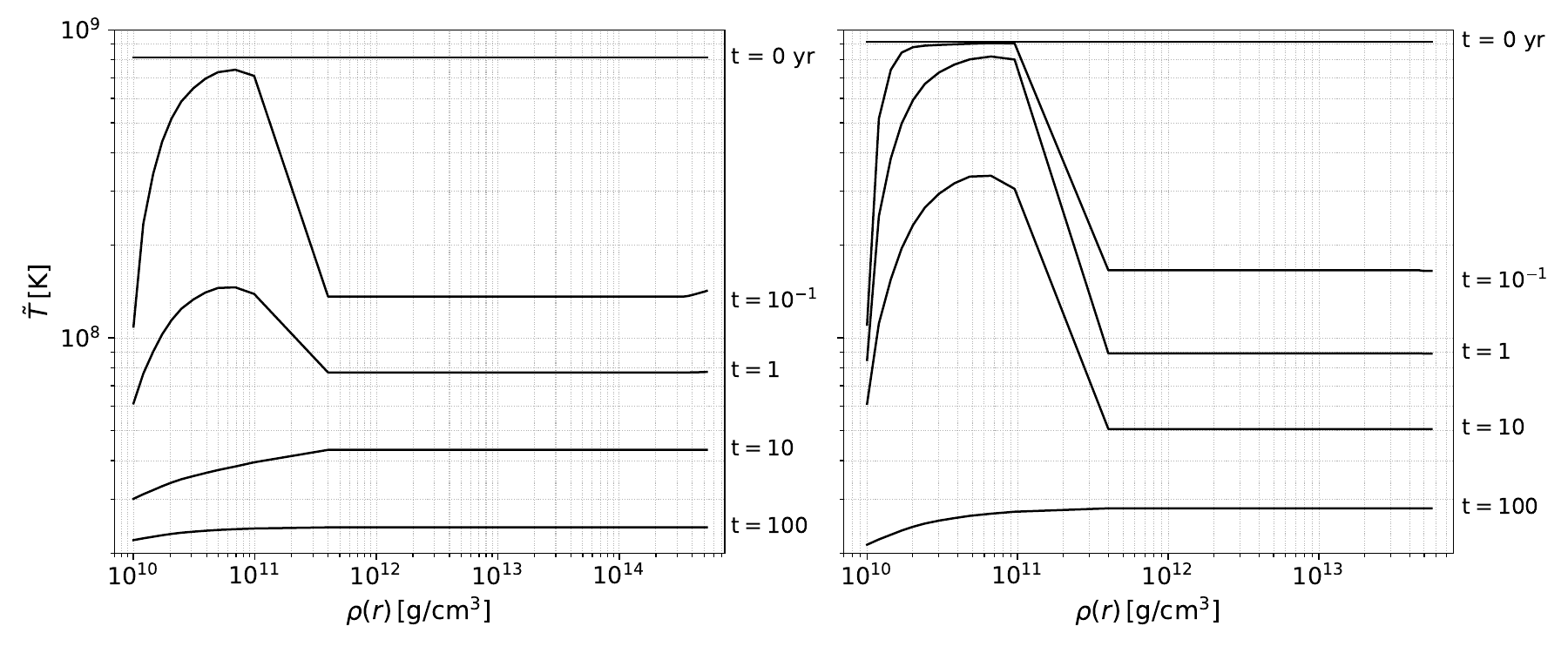}
    \end{subfigure}
    
    \vspace{1em} 

    \begin{subfigure}[b]{\linewidth}
        \centering
        \includegraphics[width=0.8\textwidth]{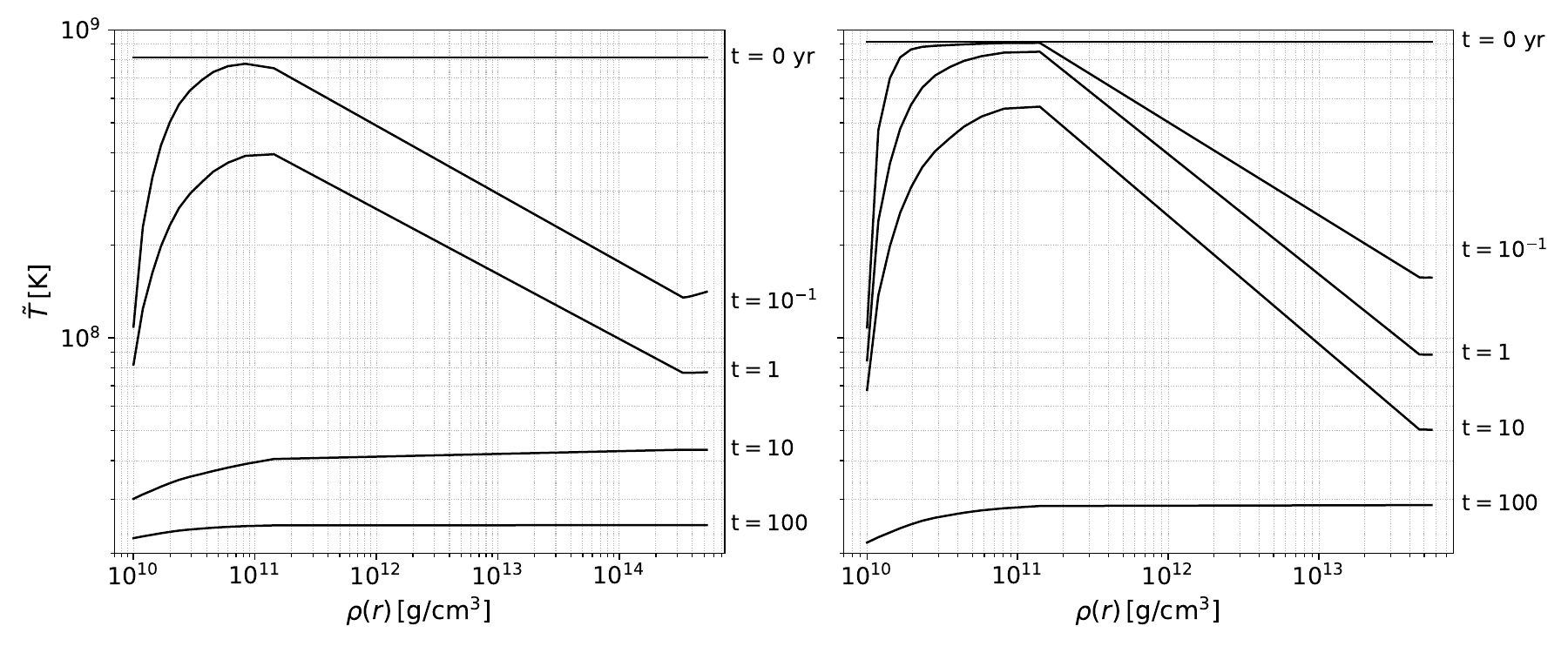}
    \end{subfigure}

    \caption{\raggedright Radial temperature profiles of a $M=1.4\ M_{\odot}$ quark star with a nuclear crust during the early thermal evolution, computed for different stellar ages $t$. Left and right columns correspond to $\bar{\Lambda}=2\mu$ and $\bar{\Lambda}=4\mu$, respectively. Top: thermal evolution computed on a radial grid that includes a point in the crust-core gap. Bottom: the gap is ignored in the numerical grid. Each curve is labeled by its corresponding age.}
    \label{fig:profile_crust}
\end{figure}

In Figure \ref{fig:profile_crust}, we show the temperature profile evolution for the two modeling approaches for a quark star with a crust. The top panels correspond to the case where the thermal gap at the crust–core boundary is taken into account, whereas the bottom panels show the results when this gap is neglected. 
The most significant differences arise during the early evolutionary stages. At these times, as shown for $t=10^{-1}$ yr, $t=1$ yr and $t=10$ yr, the temperature profile exhibits a clear discontinuity between the crust (for densities $\rho \leq4\times 10^{10}\ \mathrm{g}\ \mathrm{cm}^{-3}$) and the core ($\rho>4\times 10^{10}\ \mathrm{g}\ \mathrm{cm}^{-3}$) when we take the gap into account. The core rapidly becomes isothermal, while the crust retains higher temperatures, except in the outermost layers where surface photon emission plays an important role. 

\begin{figure}[h]
    \centering
    \includegraphics[height=7cm]{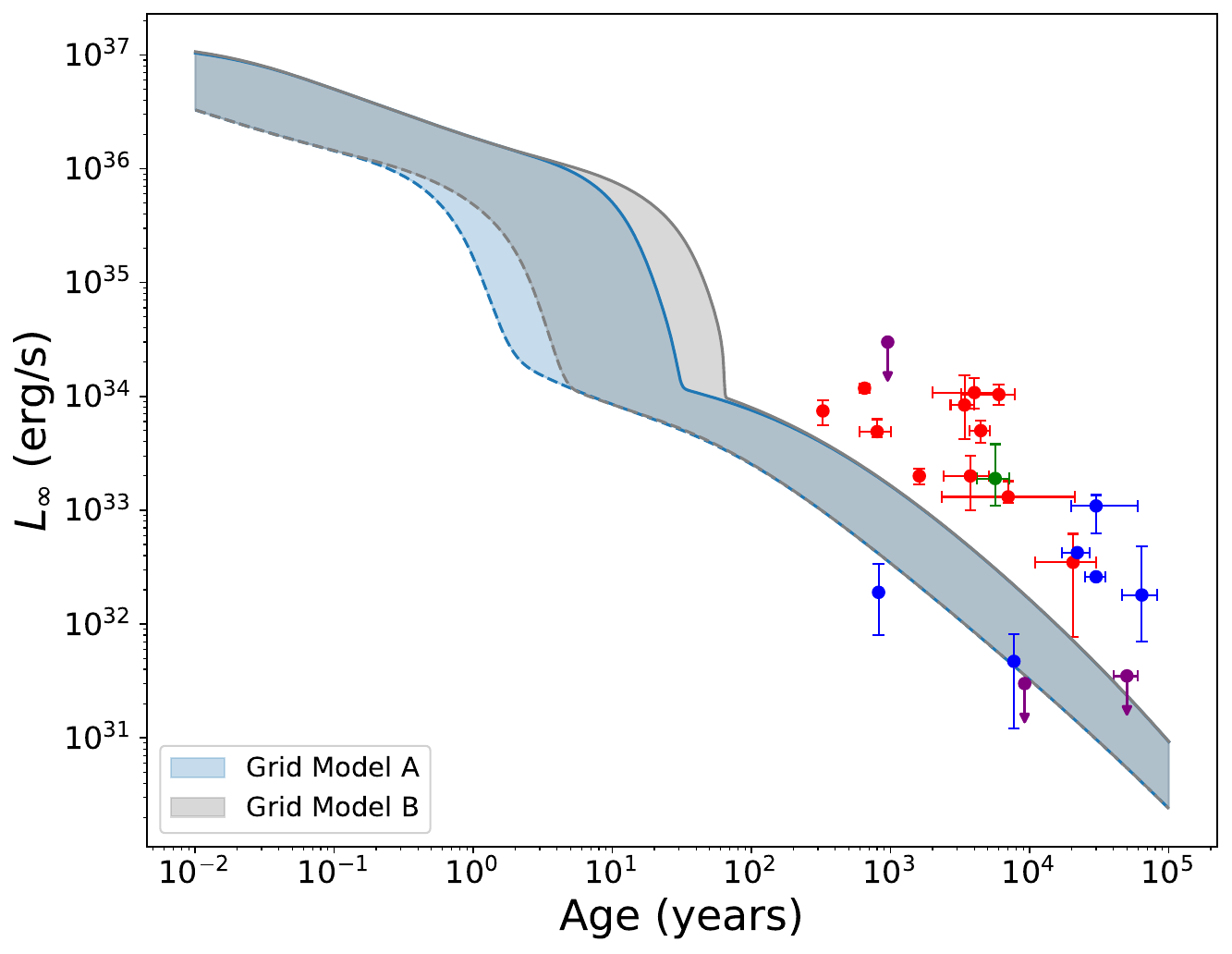}
    \caption{\raggedright Redshifted photon luminosity $L_{\infty}$ as a function of stellar age for a $M=1.4\ M_{\odot}$ quark star with a hadronic crust, comparing the cooling results obtained with Grid Model A and Grid Model B.}
    \label{fig:cooling_crust_models}
\end{figure}

In contrast, when the thermal gap is not included, the transition between crust and core is smoother, and the core takes longer to reach an isothermal state. Despite these early-time differences in the internal temperature distribution, both models converge to nearly identical thermal profiles by $t=100$ yr. As a result, the differences between the predicted luminosities appear only for ages $\lesssim 100$ yr, as illustrated in Figure \ref{fig:cooling_crust_models}. Such ages, however, are significantly younger than those of any neutron stars in our observational sample. Consequently, the early–time discrepancy between the two grid models has no practical impact on the comparison with data.

Moreover, the computational expense associated with Grid Model A is substantially higher. Accurately positioning a grid point inside the narrow crust–core interface requires resolving length scales on the order of femtometers, which makes the radial derivatives exceedingly steep and demands integration steps that are too small for the scale we need for this problem. Grid Model B avoids this difficulty while producing indistinguishable results for all relevant ages. 

Beyond the variations in the cooling band behavior, the temperature profile evolution further highlights differences between the bare and crusted quark star models. In the presence of a crust, a pronounced temperature gradient develops, with the two regions cooling at different rates. This behavior resembles that of neutron star cooling, where efficient neutrino emission in the core causes it to cool more rapidly than outer layers, leading to a ``cold wave" propagating outward toward the crust. This process delays the onset of an isothermal profile. Specifically, the time required for the temperature difference between the center ($r=0$) and the boundary ($r=R_{\rm boundary}$) to decrease to a value on the order of $10^{3}$ K is approximately $10^{4}$ yr for $\bar{\Lambda}=2\mu$ and $4\times 10^4$ yr for $\bar{\Lambda}=4\mu$.

\section{Summary and outlook}
\label{sec:summary}

We investigated the thermal emission properties of quark stars with and without a hadronic crust, using the simplest version of a pQCD equation of state for cold quark matter, up to $O(\alpha_s)$, using the $\overline{\mathrm{MS}}$ scheme, and taking into account renormalization group running effects on the coupling and the strange quark mass. 
The choice of the renormalization scale affects the stiffness of the EoS which, in turn, determines the macroscopic structure of the star. So, the cooling behavior is impacted in two distinct ways: through the macroscopic inputs determined by the solutions of the Tolman-Oppenheimer-Volkoff equations and through the microscopic inputs relevant to thermal evolution. To explore this sensitivity, which is a measure of our the uncertainty in the EoS, we consider a band in $\bar\Lambda$ from $2\mu$ to $4\mu$.

Our results show that, even when accounting for this theoretical uncertainty, bare quark stars cool too rapidly to be consistent with observational data, exhibiting luminosities lower than those of the coldest observed isolated neutron stars, as shown in Figure \ref{fig:bare_running}. We also find that the cooling band becomes significantly narrower after approximately $t=1$ yr, indicating a reduced sensitivity to the choice of renormalization scale at later times. Furthermore, we observe that increasing the stellar mass (from $1.4\ \mathrm{M}_{\odot}$ to $2\ \mathrm{M}_{\odot}$) leads to a narrower cooling band, which also results in slightly higher luminosities compared to the $1.4\ \mathrm{M}_{\odot}$ configuration. 

Motivated by these findings, we extended our analysis to quark stars with a hadronic crust. The motivation arises from the observation that the uncertainty associated with the pQCD EoS is more significant at early stages of thermal evolution, when the star has not yet achieved internal isothermality and is primarily cooling via neutrino emission. Given the known role of the crust as a thermal insulator, which delays the establishment of an isothermal interior, we investigated how the uncertainty from the pQCD EoS influences this configuration. As expected, the temperature evolution in the crusted case is less steep than in the bare configuration, bringing the model predictions into closer agreement with the observational data, as illustrated in Figure \ref{cooling_crust}. We obtained a cooling uncertainty band that lies slightly above the coldest observed young neutron stars, while still encompassing the data point of PSR B2334+61, although the full extent of its reported uncertainties is not simultaneously accommodated. 

In contrast to the bare case, the cooling band in the crusted configuration does not exhibit a significant narrowing throughout the timescale considered (up to $t=10^5$ yr), indicating a more persistent sensitivity to the EoS parameters and emphasizing the need to include equation-of-state uncertainties in cooling calculations.

We also computed the cooling curves obtained using the MIT bag model with $\mathcal{O}(\alpha_s)$ corrections. For the bare configuration, the MIT bag model predictions remain within the uncertainty band associated with the variation of $\bar{\Lambda}$ in the pQCD equation of state, except during the very early stages of the evolution. In this initial phase, the MIT bag model curve briefly falls outside the band but enters it before the star reaches an age of $1$ yr. For quark stars with a hadronic crust, however, the behavior is reversed. In this case, the MIT bag model prediction begins within the pQCD uncertainty band but subsequently departs from it, yielding higher luminosities.

Our model for the thermal evolution of quark stars described with perturbative QCD with a hadronic crust approaches the region occupied by the mentioned young neutron star observations, although it does not reproduce the full set of data. Several aspects of the microphysics may influence this discrepancy. In particular, uncertainties in the thermal properties of quark matter in the core, the treatment of the crust, and the modeling of the stellar envelope may significantly affect the location of the predicted cooling band relative to the observations. Furthermore, there are additional stellar properties and effects not considered in our current framework, which we discuss below.

It is well established that color superconductivity can modify both the neutrino emissivity and the heat capacity of quark matter \cite{Shovkovy:2002gx, Blaschke:2005dc, Noda:2011ag, Schmitt:2017efp}. A natural extension of the present study would therefore be to investigate the cooling of quark stars in different color-superconducting phases, while consistently incorporating the uncertainties from the pQCD equation of state. Previous studies have shown that the presence of a color–flavor locked (CFL) phase \cite{Zapata_2022} can significantly modify the early thermal evolution of strange quark stars with a nuclear BPS crust. The strong suppression of neutrino emission in this superconducting phase can alter the predicted cooling curves, particularly at early ages, and may therefore influence how theoretical cooling bands compare with the available data.

Another important aspect concerns the role of magnetic fields. Although we compare our results with observations of high-B pulsars, magnetic-field effects are not explicitly included in our cooling calculations. Significant progress has been made in this area, such as the study by \cite{Shovkovy:2026aci}, which investigates magnetic field effects on the neutrino emission of ungapped quark matter, and the work by \cite{Negreiros:2025xpe}, which analyzes the axisymmetric cooling of neutron stars.

Regarding neutrino emissivity, although the presence of a strong magnetic field opens the channel for neutrino-antineutrino synchrotron emission, its rate remains several orders of magnitude smaller than that of the direct Urca process. For this dominant process, it has been shown that, on average, the energy emission rate decreases with increasing field strength, being suppressed by about $20\%$, even for magnetic fields as strong as $10^{17}$ Gauss. As a result, the authors do not expect this to cause a substantial difference in the overall cooling behavior. An interesting exception to this general trend arises in the extreme lowest Landau level (LLL) regime, where the emission rate can actually be significantly enhanced. Furthermore, it is worth noting that the complex interplay between color-superconducting phases and strong magnetic fields remains far from fully understood.

In addition to altering local weak-interaction rates, intense magnetic fields break the spherical symmetry of the star, introducing further complexities into the cooling process. Recent axisymmetric analyses of neutron star cooling have investigated this effect. Of particular interest is the effect of high magnetic fields on the gravitational mass observed at infinity. When comparing two stars with the same gravitational mass, a highly magnetized star will have a lower baryonic content and, consequently, a lower central density. In standard neutron stars, this density reduction directly restricts the core volume where the direct Urca process is allowed to operate. Even though the kinematic conditions for the direct Urca process are much more relaxed in quark stars, it was shown that strong magnetic fields inherently modify how heat is transported within the stellar interior and alter the thickness of the crust \cite{PhysRevD.96.123005}, non-linearly increasing the thermal relaxation time of the star. While such treatments have been explored in the nucleonic context, the impact of these changes on the cooling of quark stars has not yet been systematically investigated.

Another relevant direction is to refine the microphysical inputs by implementing up-to-date calculations of neutrino emissivity \cite{Schafer:2004jp, Adhya:2013maa} and heat capacity in quark matter \cite{Gerhold:2004tb}.

\begin{acknowledgments}
We thank A. Schmitt for useful discussions. This work was partially supported by INCT-FNA (Process No. 464898/2014-5), CAPES (Finance Code 001), CNPq, and FAPERJ.
\end{acknowledgments}



\bibliographystyle{apsrev4-1}
\bibliography{references.bib}

\end{document}